\documentstyle[twoside,epsfig]{actaps}
\def\day{18 September 1998}
\def\autor{A. Di Giacomo}
\def\shorttitle{EXPLORING ENSEMBLES OF INSTANTONS.}

\begin{document}

\title{{EXPLORING ENSEMBLES OF INSTANTONS.}}
\author{A. Di Giacomo\email{digiacomo@pi.infn.it}}{
Dip. Fisica Universit\`a and INFN, Via Buonarroti 2 ed.B 56126
PISA, ITALY}

\author{Dedicated to Jan Pisut on the occasion of his 60th birthday.}{}


\abstract{A number of observables is constructed which can give
useful information  on instanton ensembles.
The basic properties used are (1) Istantons are $SU(2)$ configurations. (2) They
are self-dual or antiself-dual.
}

\section{Introduction}
The discovery of instantons\cite{1} first opened a way to investigate non
perturbative effects in QCD\cite{2}.

Multi instanton-antiinstanton configurations are not solutions of the
equations of motion: they are approximate solutions in the limit in which their
average distances are large compared to their average size (dilute gas regime).

The early approaches to instanton physics\cite{3} were based on the assumption
that QCD dynamics would choose such a regime, thus providing a key to non
perturbative effects in hadron physics.
It was rapidly realized, however, that this was not the case. Instantons do
overlap and interact in the QCD vacuum, and are more like a liquid than as a
rarefied gas\cite{2}. It was also understood, however, that instantons are
relevant configurations, especially in connection with chiral symmetry. For this
reason models were developed to describe  multiinstanton
configurations\cite{4,5,2} and their probability distribution (instanton
ensembles). Usually these models are tested by their predictions for different
physical quantities, or by comparison to lattice configurations.

Lattice configurations can be polished of the quantum fluctuations by
cooling techniques, and their semiclassical background can be
exposed\cite{6,7}. Improved techniques allow to reduce deformations of the
instantons during the cooling process \cite{8}.

Different model ensembles put emphasis on different physical aspects. They can
look equally good for  a number of predictions, differ in others, and it is not
easy to discriminate between them. It would be useful to identify  a number of
observables which can be determined, e.g. on the lattice and on model ensembles,
and which can provide a test of model ensembles.

This paper is a contribution in that direction.
We suggest a number of observables, which can be sensitive to correlations at
different distances or to size distribution of the instantons.

We will make use of two basic ingredients
\begin{itemize}
\item[(i)] Instantons are $SU(2)$ configurations.
\item[(ii)] The gauge field of an (anti)instanton is (anti)self dual, or
\begin{equation}
\vec E^a = \pm\vec B^a \label{eq:1}\end{equation}
\end{itemize}
$\vec E^a$, $\vec B^a$ are the a-th colour component of the chromoelectric and
chromomagnetic fields.

In sect.2 we shall elaborate on some consequences of the property (i) above, in
sect.3 on some consequences of the property (ii).

Sect.4 will contain some concluding remarks.

\section{Instantons as $SU(2)$ configurations.}
The fact that instantons, as configurations belonging to an $SU(2)$ subgroup of
colour $SU(3)$, have specific symmetry properties, was exploited in ref\cite{9}
to derive selection rules in physical processes.

We first discuss how the $SU(2)$ nature of a configuration can be
detected. For the sake of definiteness we will have in mind lattice
configurations, and hence parallel transports, i.e. elements of the gauge
group. It is immediate to translate the discussion in the language of the
elements of the Lie algebra, like the gauge fields in the continuum.

Let $U$ be any element of the gauge group in the fundamental representation,
which can be a link or any parallel transporter. A parallel transport along a
closed path, in particular, has gauge invariant trace.

$U$ can be  diagonalized by a unitary transformation, obtaining
\begin{equation}
U_D = \left(\matrix{ e^{i \alpha} & 0&0\cr 0&e^{i\beta}&0\cr 0 & 0&
e^{-i(\alpha+\beta)}\cr}\right) \label{eq:2}\end{equation}
One of the elements is fixed in terms of the other two by the condition that
$\det U_D = \det U = 1$. From eq.(\ref{eq:2}) we have
\begin{equation}
{\rm Tr} U = {\rm Tr} U_D =
\cos\alpha+\cos\beta + \cos(\alpha+\beta) + i\left(
\sin\alpha+\sin\beta - \sin(\alpha+\beta)\right) \label{eq:3}\end{equation}
The necessary and sufficient condition for $U$ to be an element of $SU(2)$ is
that one of the diagonal elements of $U_D$, eq.(\ref{eq:2}) is equal to 1, or
that either $\alpha$, or $\beta$, or $(\alpha+\beta)$ vanishes. This is also
the necessary and sufficient condition for ${\rm Tr} U$ to be real
\begin{equation}
Im {\rm Tr} U = 0 \label{eq:4}\end{equation}
If $U$ is a parallel transporter inside a classical instanton configuration
eq.(\ref{eq:4}) is satisfied.

Quantum fluctuations will in general spoil this property, and the same will do
the overlapping of different instantons.

If $U$ is a link on the lattice, the measure $d U$ of the Feynman path integral
induces a distribution of values in the complex plane $Re[Tr(U)]$, $Im[Tr(U)]$. Let us
define polar coordinates $\rho$, $\theta$ in that plane
\[\rho = \sqrt{(Re[Tr(U)])^2 + (Im[Tr(U)])^2}\qquad \theta =
{\rm arctg}\frac{Im[Tr(U)]}{Re[Tr(U)]}\]
Then the allowed region in defined by the condition
\begin{equation} f^2(\rho,\theta) =
27 - \rho^4 - 18 \rho^2 + 8 \rho^3\cos 3\theta \geq 0 \label{eq:5}\end{equation}
and is the region inside the curved triangle in fig.1.
\par\noindent
\begin{figure}[t]
\begin{center}\mbox{\input epsf \epsfysize 2.2in
                        \epsfbox{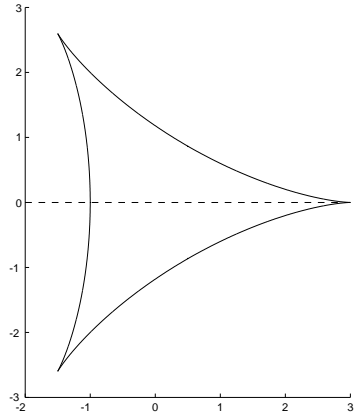}}\end{center}
\caption{Fig. 1. The allowed region in the complex plane for $Tr(U)$,
$U\in SU(3)$. Real axis corresponds to $SU(2)$ elements.}
\end{figure}
\par\noindent
The distribution is
\begin{equation}
P(\rho,\theta) d \rho d \theta = |f(\rho,\theta)| \frac{\rho d \rho d \theta}{
2 \pi^2}\label{eq:6}\end{equation}
At $\beta = 0$ the Feynman path integral
\begin{equation}
Z(\beta) = \int\prod_{\mu n} d U_\mu(n) \exp(-\beta S)
\label{eq:7}\end{equation}
reduces to
\begin{equation}
Z(\beta=0) = \int\prod_{\mu n} d U_\mu(n)
\label{eq:8}\end{equation}
and the trace of any parallel transport is distributed according to
eq.(\ref{eq:6}).

At finite $\beta$ the distribution will be different, it will in general depend
on the path which identifies $U$, and will be non zero out of the real axis in
the allowed region of fig.(1).

If the configuration is cooled and one single instanton or a very dilute gas of
instantons is left, then eq.(\ref{eq:4}) will be valid and the distribution in
the complex plane of fig.1 will shrink to a thin line along the real axis, at
least for paths of the parallel transporter $U$ with size smaller than the
average size of instantons. For paths comparable or larger than the size a non
zero immaginary part is allowed.

A systematic study with paths of different size could then give information on
the distribution of instanton sizes. If the gas is not dilute the distribution
along the imaginary axis will shrink by cooling, but will not go to zero.

Preliminary exploration on the lattice of the effect of cooling show that
instantons strongly overlap\cite{10}.

If $U$ is taken to be a plaquette, then at high $\beta$ values of $U$ will be
most probably be
near to the unit matrix, i.e. it will be $\rho\simeq 3$, which is
the vertex on the right of fig.1, and the density of action will be approximately
\[ S = \left[1 - \frac{1}{3} Re[Tr(U)]\right] \simeq 1 - \frac{\rho}{3}\]
In a cooled configuration regions with higher values of $S$, like the central
regions of instantons or antiinstantons, can be selected by cutting out the
higher values of $S$ or the lower values of $Re[Tr(U)]$.
One can e.g. look at slices of the distribution
in fig.1
parallel to the imaginary axis, at different values of $Re[Tr(U)]$.
Looking at the central region of instantons
should give a distribution more concentrated at $Im[Tr(U)]=0$.
This effect has  been observed\cite{10}.

The transport around a bigger loop than a plaquette, (e.g.
$2\times2$ plaquette) should show a similar distribution. Again when the size of
the loop becomes larger than the size of the instanton the effect should
disappear.

More generally a systematic study of the distribution with respect to
$Im[Tr(U)]$, for different choices of $U$, on ensembles and on lattice can
provide information on the size distribution and on the distribution in space
time of instantons.
\section{Self duality.}
Consider the quantity
\begin{equation}
{\cal Q} = G_{\mu\nu} G^*_{\mu\nu} \label{eq:9}\end{equation}
with $G_{\mu\nu} = G^a_{\mu\nu} T^a$, $G^*_{\mu\nu} = \frac{1}{2}
\varepsilon_{\mu\nu\rho\sigma} G_{\rho\sigma}$, and $T^a$ the generators in the
fundamental representation. On a classical configuration
\begin{equation}
{\cal Q} = 2 \vec E\cdot\vec H = 2 \vec E^a\vec H^b T^a
T^b\label{eq:10}\end{equation}
In general
\[ T^a T^b = \frac{1}{2}[ T^a , T^b] + \frac{1}{2}\{ T^a , T^b\}\]
and since
\begin{equation}
[ T^a , T^b] = i f^{abc} T^c\qquad
\frac{1}{2}\{ T^a , T^b\} = \frac{1}{2 N_c}\delta^{ab} + d^{abc} T^c
\label{eq:11}\end{equation}
\begin{equation}
{\cal Q} = \vec E^a\vec H^b\left\{
i f^{abc} T^c + \frac{1}{2 N_c}\delta^{ab} + d^{abc} T^c\right\}
\label{eq:12}\end{equation}
For a self dual configuration, by use of eq.(\ref{eq:1}) the term in $f$
vanishes. If in addition the configuration is $SU(2)$ also the term in $d$
vanishes and
\begin{equation}
{\cal Q} = \frac{\vec E^a\vec H^a}{N_c} = \frac{{\rm Tr} {\cal Q}}{N_c}
\label{eq:13}\end{equation}
i.e. ${\cal Q}$ is a singlet, proportional to the topological charge density.
In general
\begin{equation}
{\cal Q}^c = 2 {\rm Tr}T^c {\cal Q} =
\vec E^a\vec H^b\left\{ i f^{abc} + d^{abc}\right\}
\label{eq:14}\end{equation}
The imaginary part of ${\cal Q}^c$ reflects the deviations from self duality,
the real part the deviations from $SU(2)$. On an instanton configuration, or on
a dilute gas, they are separately zero, site by site.

On the lattice the operator
\begin{equation}
\tilde {\cal Q}(x) =
\frac{1}{2i}\left[
(\Pi_{\mu\nu} - \Pi^{\dagger}_{\mu\nu})
(\Pi^*_{\mu\nu} - \Pi^{*\dagger}_{\mu\nu})\right] \label{eq:15}\end{equation}
with
\[ \Pi^*_{\mu\nu} = \frac{1}{2}\varepsilon_{\mu\nu\rho\sigma}\Pi_{\rho\sigma}
\]
is a covariant operator and transforms as an octet. In the continuum
limit it is
proportional to the operator ${\cal Q}$.

For self dual or antiselfdual fields, where $\Pi_{\mu\nu} =\Pi^*_{\mu\nu}$,
or $\Pi^\dagger_{\mu\nu}$
it vanishes identically.
The correlator
\begin{equation}
\langle 0|{\rm Tr}\left[ \tilde{\cal Q}(x) U(x,0) \tilde{\cal Q}(0)
U^\dagger(x,0)\right]|0\rangle
\label{eq:16}\end{equation}
is rigoroursly zero on single instantons or dilute gas. On cooled
configurations it can explore the overlap of instantons and antiinstantons at
small distances.

Other correlators can be constructed on the same lines. They should however be
tested on ensembles to fully understand their significance.

\section{Concluding remarks.}
Identifying and studying a number of observables which can provide information on
the distribution and interaction of the background instantons in QCD is a
theoretically relevant program. The idea of this paper is to use the two
specific properties of instantons, their $SU(2)$ nature and self duality as a
key in this research.

We have made a few suggestions which can be developed in many directions, by
inventing new operators. Further work of checks and of comparison with
numerical data is needed, but we think that exploiting the above properties of
instanton is the correct strategy.

\section*{Acknowledgements}
Partially supported by EC TMR Program
ERBFMRX-CT97-0122, and by MURST, project: ``Fisica Teorica delle
Interazioni Fondamentali''.


\section*{References}
\begin{thebibliography}{99}
\bibitem{1} A. Belavin, A. Polyakov, A. Schwartz, Yu. Tyupkin,
{\em Phys. Lett.} B {\bf 59}, 85 (1975).
\bibitem{2} For a recent review of instantons see:
T. Sch\"afer, E.V. Shuryak, {\em Rev. Mod. Phys.} {\bf 70 }, 323 (1998) and
{\em hep-ph 9610451}.
\bibitem{3}C.G. Callan, R.F. Dashen, D.J. Gross, {\em Phys. Rev.} D {\bf 17},
2717 (1978), {\em Phys. Rev.} D {\bf 18}, 4684 (1978),
{\em Phys. Rev.} D {\bf 19}, 1826 (1979).
\bibitem{4}E.V. Shuryak, {\em Nucl. Phys.} B {\bf 302} 559, 574, 599 (1988).
\bibitem{5}D.I. Dyakonov, V.Y. Petrov,
{\em Nucl. Phys.} B {\bf 272} 457 (xxxx).
\bibitem{6}M.C. Chu, J.M. Grandy, S. Huang, J.W. Negele,
{\em Phys. Rev.} D {\bf 49}, 6039 (1994).
\bibitem{7}C. Michael, P.S. Spencer,
{\em Phys. Rev.} D {\bf 50}, 7570 (1994).
\bibitem{8}M. Garcia Perez, A. Gonzales Arroyo, J. Snippe, P. van Baal,
{\em Nucl. Phys.} B {\bf 413} 535 (1994).
\bibitem{9}B.V. Geshkenbein, B.L. Ioffe,
{\em Nucl. Phys.} B {\bf 166} 340 (1980).
\bibitem{10}L. Vergnano, {\em Tesi di laurea}, Pisa 1994.
\bibitem{11}A. Di Giacomo, H. Panagopoulos, L. Vergnano,
(1993) unpublished.
\end{thebibliography}
\end{document}